\documentclass[
    ,final            % use final for the camera ready runs
  ]
  {aipproc}
\usepackage{wrapfig}
\layoutstyle{8x11single}
\begin{document}

\title{Bottomonium at Non-zero Temperature from Lattice Non-relativistic QCD}

%11.15.Ha,12.38.Gc,14.40.Pq}
%<Replace this text with PACS numbers; choose from this list:
%                \texttt{http://www.aip..org/pacs/index.html}>}
\author{Gert~Aarts}{
  address={Department of Physics, Swansea University, Swansea SA2 8PP, 
United Kingdom}
}
\author{Chris~Allton}{
  address={Department of Physics, Swansea University, Swansea SA2 8PP, 
United Kingdom}}
\author{Seyong~Kim}{
  address={Department of Physics, Sejong University, Seoul 143-747, Korea}
}
\author{Maria Paola~Lombardo}{
  address={INFN-LNF, I-00044, Frascati (RM) Italy, Humboldt-Universit\"at zu Berlin, 12489 Berlin,Germany}
}

\author{Mehmet~B.~Oktay}{
  address={Physics Department, University of Utah, Salt Lake City,
  Utah, USA}
}
\author{Sinead~M.~Ryan}{
  address={School of Mathematics, Trinity College, Dublin 2, Ireland}
}
\author{D.~K.~Sinclair}{
  address={HEP Division, Argonne National Laboratory, Argonne, IL 60439, USA}
}
\author{Jon-Ivar~Skullerud}{
  address={Department of Mathematical Physics, National University
  of Ireland, Maynooth, County Kildare, Ireland}
}
\begin{abstract}
The temperature dependence of bottomonium states at temperatures above and 
below $T_c$ is presented, using non-relativistic dynamics for the bottom quark and 
full relativistic lattice QCD simulations for two light flavors on a highly 
anisotropic lattice. We find that the S-waves ($\Upsilon$ and $\eta_b$) show little  
temperature dependence in this range while the P wave propagators show a crossover from 
the exponential decay characterizing the hadronic phase to a power-law 
behavior consistent with nearly-free dynamics at approximately twice the 
critical temperature. 
\end{abstract}
\maketitle

%%%%%%%%%%%%%%%%%%%%%%%%%%%%%%%%%%%%%%%%%%%%
%% MAINMATTER
%%%%%%%%%%%%%%%%%%%%%%%%%%%%%%%%%%%%%%%%%%%%
\subsection{Introduction}
Heavy quark states can be an important probe of the dynamics of the Quark Gluon Plasma (QGP). 
While charmonium suppression~\cite{Matsui:1986dk} has been observed at a range of 
energies, a number of sometimes competing effects complicate the interpretation of these 
patterns. Many of these difficulties are not present for bottomonium making suppression 
patterns observed in that regime in principle more straightforward to interpret. 
Intriguing early results from experiments at CMS~\cite{cms_may05} and 
STAR~\cite{rosi_reed} demonstrating suppression in the Upsilon system suggest a lattice 
calculation is timely and may help to shed light on the link to the spectrum of 
bound states. 

In this paper we use dynamical anisotropic lattice gauge configurations and treat the 
bottom quarks using non-relativistic QCD (NRQCD) 
at a range of temperatures, $0.42T_c < T< 2.09T_c$. 
NRQCD has been used extensively to study 
heavy quark physics both on and off the lattice. 
At finite temperature this formalism offers a number of technical advantages over the 
relativistic theory.  Firstly, the temperature dependence in 
NRQCD correlators is due only to the thermal medium, with no contribution from 
thermal boundary conditions.  Secondly, there is no nontrivial spectral 
weight at zero energy, which would yield a constant time-independent contribution to 
the correlators complicating the analysis~\cite{kernal} and casting 
doubt on the results for melting or survival of charmonium at high 
temperatures~\cite{Umeda:2007hy}.
 
The non-relativistic theory is also the first effective theory to be obtained when integrating out 
the UV degrees of freedom~\cite{Brambilla:2010vq,Burnier:2007qm} and requires only 
that the heavy quark mass, $M\gg T$ which is reasonable 
for $b$-quarks at the temperatures in our simulations, up to $2T_c\sim400$ MeV. 
A more detailed discussion is in Ref.~\cite{Aarts:2010ek} and 
references therein where some results shown 
here have also appeared. A complete MEM analysis will be presented in a forthcoming paper.  
\subsection{Theoretical formalism and results}
In this paper we present results from a study of the temperature dependence of S and P 
waves (in the $\Upsilon$ and $\chi_{b1}$ channels respectively). 
From Ref.~\cite{Burnier:2007qm} the correlators for S and P states, in continuum NRQCD with $E_p=p^2/2M$ and in the absence of interactions, are  
\begin{equation}
G_{S}(\tau)  \sim \int \frac{d^3p}{(2\pi)^3}\, \exp(-2E_{\mathbf{p}}\tau) \sim \tau^{-3/2}, 
\;\;\;\;{\rm and} \;\;\;\;
G_{P}(\tau)  \sim \int \frac{d^3p}{(2\pi)^3}\, \mathbf{p}^2 \exp(-2E_{\mathbf{p}}\tau)  \sim \tau^{-5/2},
\end{equation}
indicating a power-law decay at large euclidean time. Although in a lattice simulation 
this behaviour can be modified by interactions and 
lattice artifacts it provides a useful guide for what to expect. 

In this paper we also present preliminary results from a maximum entropy 
method (MEM)~\cite{Asakawa:2000tr} analysis of bottomonium S waves. 
The dataset used in the MEM analysis is modified 
slightly - including more configurations and a larger range of temperatures as well as a more 
accurate nonperturbative tuning of the bare anisotropy, $\xi=a_s/a_\tau$ in the action. 
The gauge configurations are those used in Ref.~\cite{Oktay:2010tf} with 
$\xi=a_s/a_\tau = 6$. NRQCD bottomonium (point) propagators, using an action including terms up to 
$\mathcal{O}(v^4)$, were computed and analysed for a number of S and P wave channels. 
\begin{table}[h]
\begin{tabular}{llllll}
\hline
\multicolumn{1}{c}{$N_s$\hspace{0.5cm}}    & 
\multicolumn{1}{c}{$N_\tau$\hspace{0.5cm}} & 
\multicolumn{1}{c}{$T$(MeV) [in Ref~\cite{Aarts:2010ek}]} &
\multicolumn{1}{c}{$T/T_c$}                & 
\multicolumn{1}{c}{No.\ of Configurations} & 
\multicolumn{1}{c}{No. of Configurations used in Ref~\cite{Aarts:2010ek}]} \\ 
\hline 
12 & 80 &  90         & 0.42 &  250 & 74 \\      
12 & 32 &  221        & 1.05 & 1000 & 500 \\ 
12 & 28 &  263        & 1.20 & 1000 &     \\
12 & 24 &  294 [306]  & 1.40 & 500  & 500 \\ 
12 & 20 &  368        & 1.68 & 1000 &     \\
12 & 18 &  408        & 1.86 & 1000 &     \\
12 & 16 &  441 [458]  & 2.09 & 1000 & 500 \\ 
\hline
\end{tabular}
\caption{Some lattice details for this work. 
The lattice spacing is $a_\tau^{-1}\approx 7.35$ GeV, 
set using the 1P-1S spin-averaged splitting in charmonium~\cite{Aarts:2007pk}.}
\label{tab:lattdetails}
\end{table}
The analysis presented in~\cite{Aarts:2010ek} and described briefly below was based on a subset of the 
temperatures now available (as indicated in the table). 
Full details of the nonperturbative tuning of the anisotropy are in 
Refs.~\cite{Aarts:2007pk,Morrin:2006tf}. 
\subsubsection{Temperature dependence of the $\Upsilon$ and $\chi_b$ states}
To distinguish bound and free states we consider the effective mass, 
$m_{\rm{eff}}(\tau)= -\log\frac{G(\tau)}{G(\tau-a_\tau)}$ and define a new quantity: 
the effective power, $\gamma_{\rm{eff}}(\tau) = -\tau\frac{G(\tau+a_\tau)-
G(\tau -a_\tau)}{2a_\tau G(\tau)}$. 
\begin{figure}[h]
  \includegraphics*[width=0.43\linewidth]{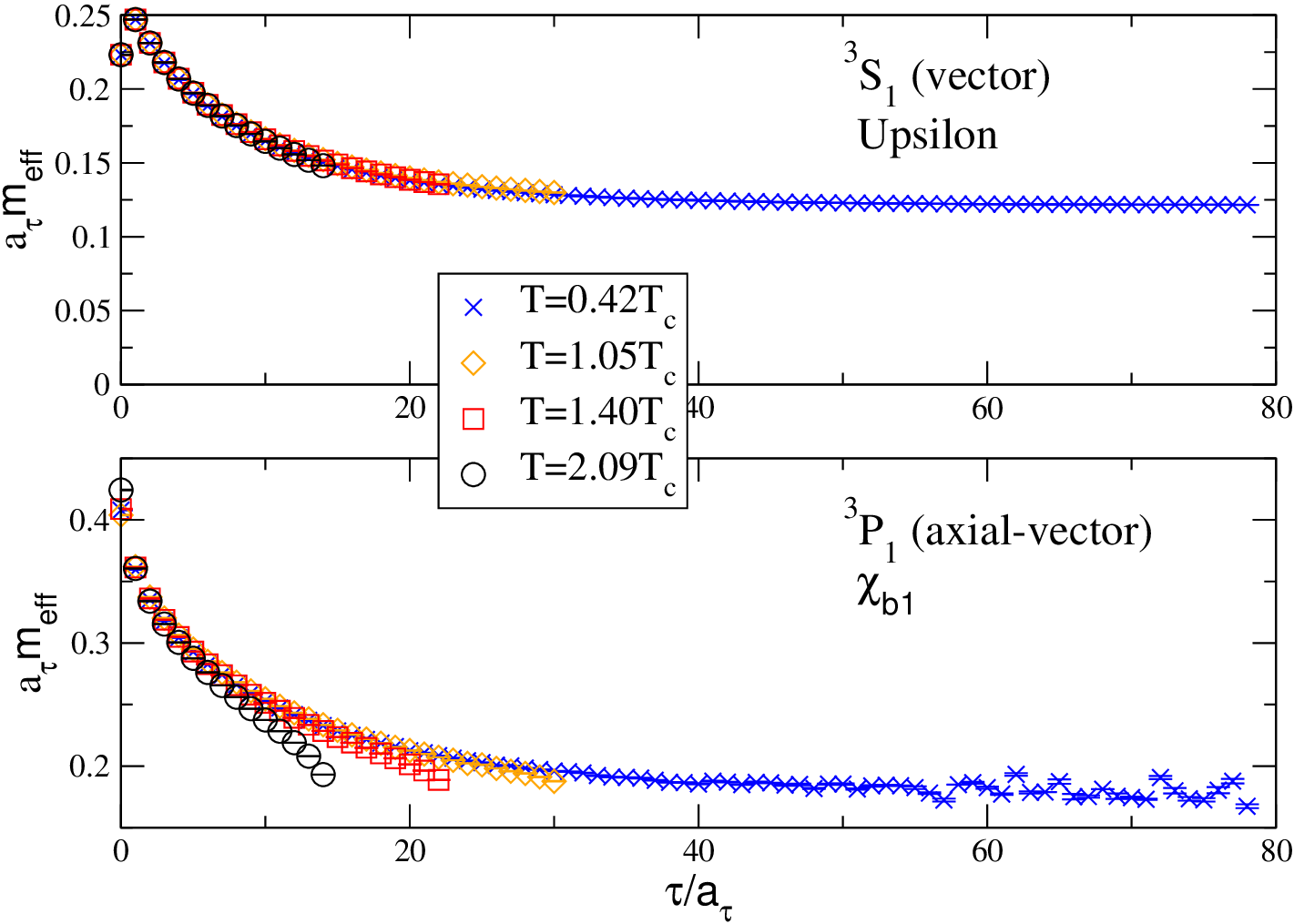}
  \includegraphics*[width=0.43\linewidth]{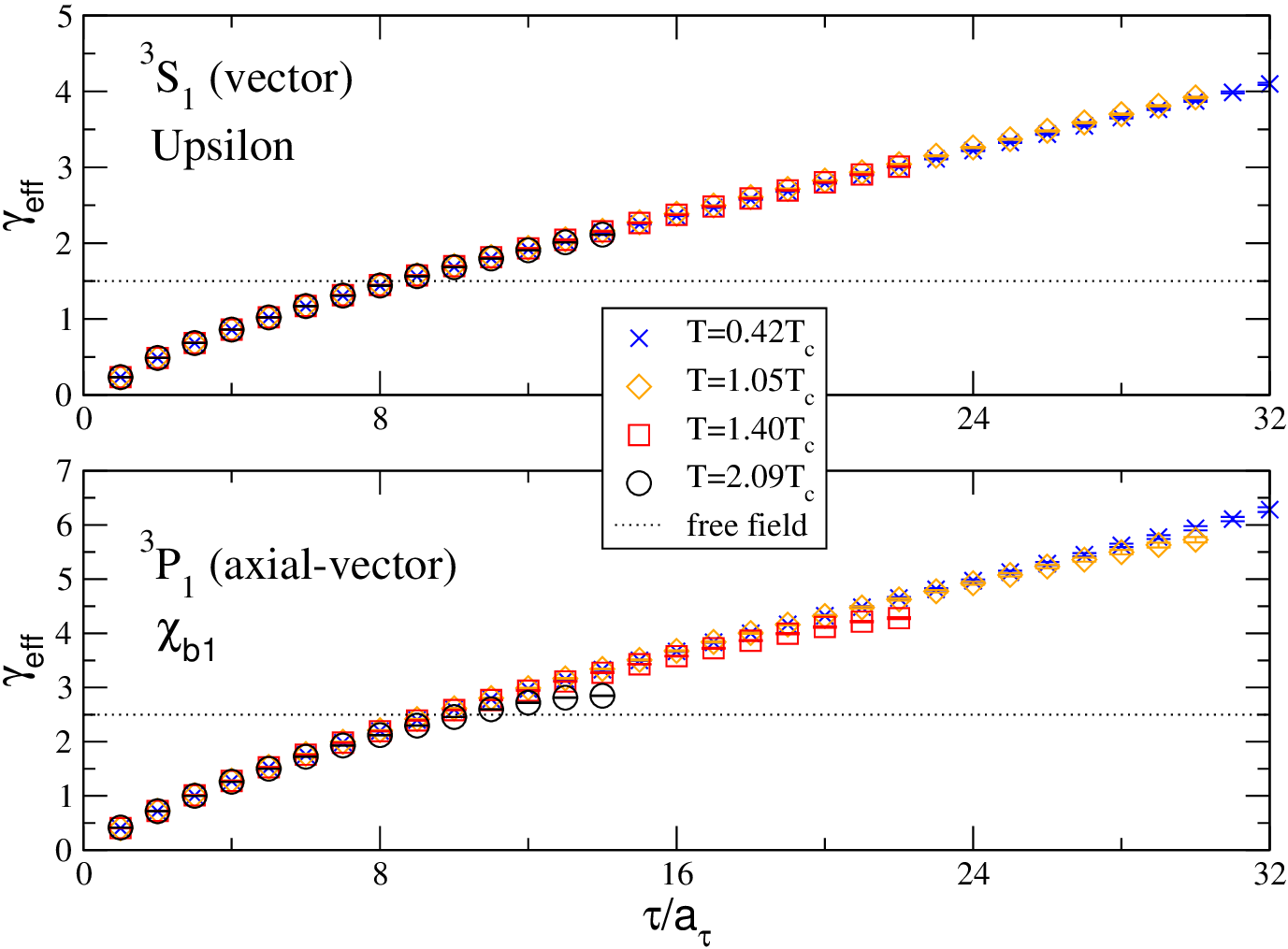}
  \caption{The left panel shows the effective mass for the S and P wave states. 
The right panel is the effective power for the same states. Each pane shows a range of 
temperatures.}
  \label{fig:effmass_power}
\end{figure}
For states in the hadronic phase we expect 
to see the exponential decay characteristic of bound states, with 
$m_{\rm{eff}}(\tau)$ becoming a constant at large $\tau$. When quarks are 
unbound we expect the correlators will show a power-law behavior and that 
$\gamma_{\rm{eff}}(\tau)$ will tend towards a constant. These quantities are shown in 
Fig~\ref{fig:effmass_power}.
We note that the P wave behaviour at $T>T_c$ rules out pure exponential decay, 
which we interpret as a medium modification at temperatures above $T_c$. In addition, we 
see a tendency of the P wave to flatten out and approach the free continuum behavior at 
large $T$ ($\sim 2T_c$). 
\paragraph{Maximum entropy analysis}
To study in detail the temperature 
dependence of the bottomonium correlators we calculate the spectral functions with MEM 
using the relation,  
$\displaystyle{G(\tau) = 
\int_{-2M}^\infty\frac{d\omega'}{2\pi}\, e^{-\omega'\tau} \rho(\omega')\,}$, where 
$\omega=2M+\omega'$ and dropping terms suppressed when 
$M\gg T$~\cite{Burnier:2007qm}. 
\begin{figure}[h]
\centering
  \includegraphics*[width=0.4\textwidth]{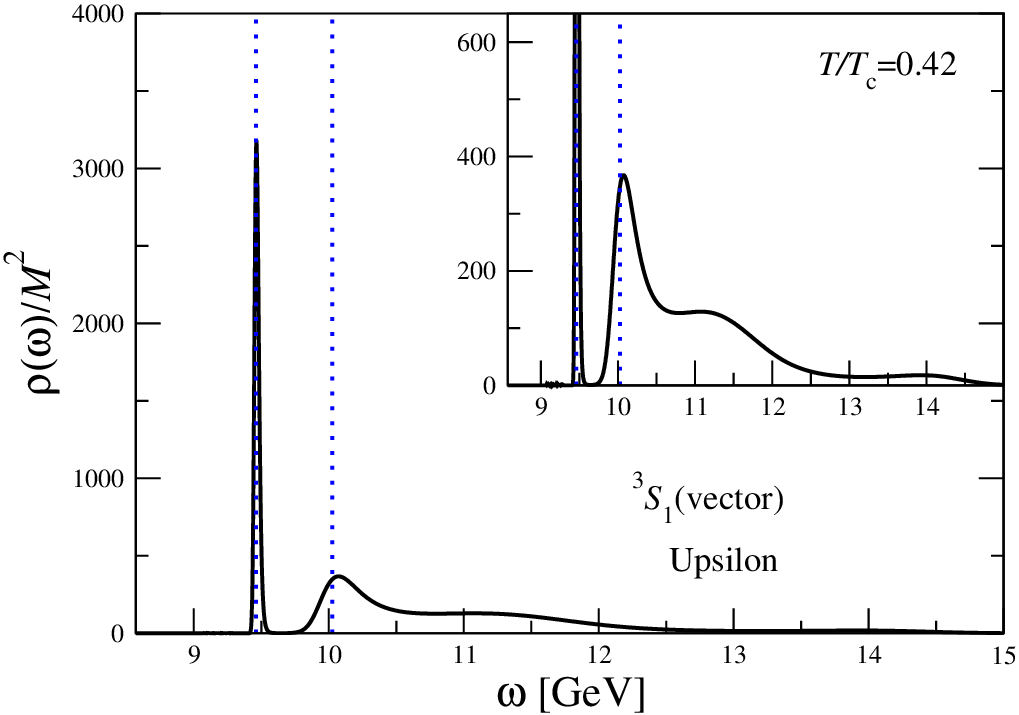}
  \includegraphics*[width=0.49\textwidth]{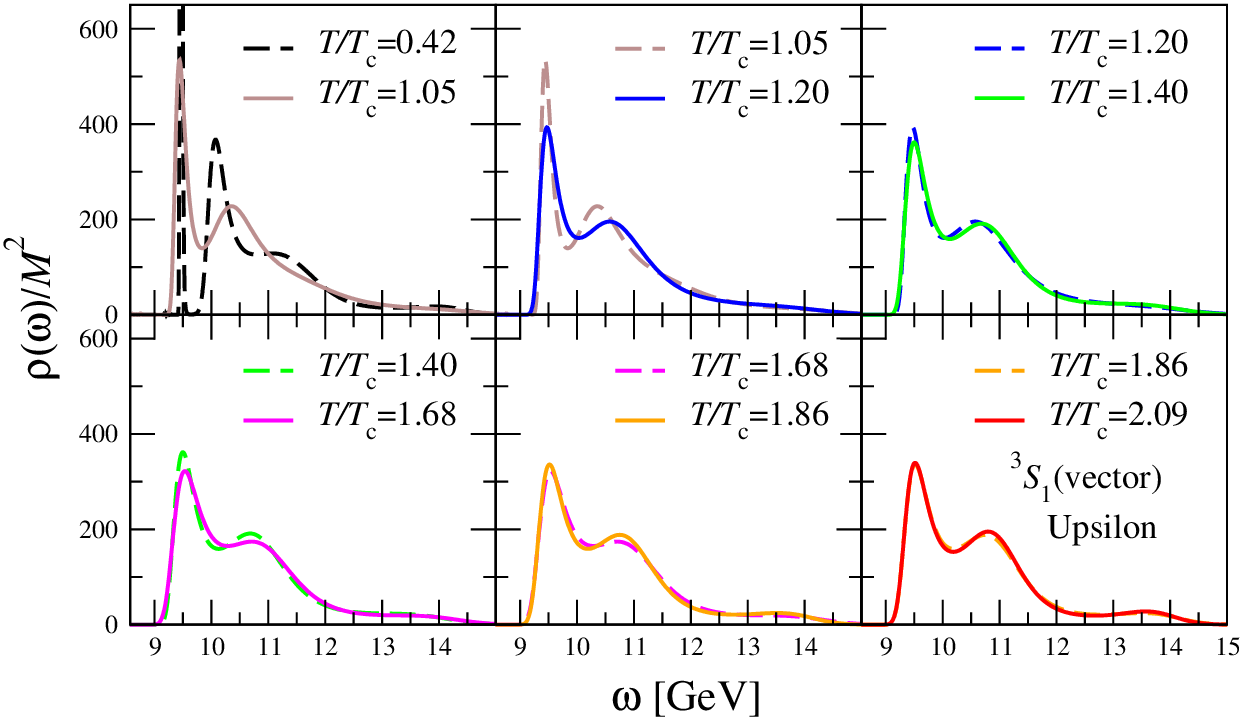}
  \caption{$\Upsilon$ spectral functions, normalised by the heavy quark mass; at zero temperature (left) and nonzero temperatures (right).} 
  \label{fig:zerotemp_mem}
\end{figure}
For the MEM to give reliable results a sufficient number of points in the euclidean time direction is 
required: ${\cal O}(10)$ independent lattice points is a good guide. At temperatures 
close to $2T_c$ this would necessitate a lattice spacing $a\sim 0.025$ fm, which on an isotropic lattice is prohibitively costly. 
The anisotropic lattice provides an excellent solution: with $a_\tau\ll a_s$ a sufficient number of points for MEM to work can be simulated for reasonable numerical cost. The MEM analysis is performed using Bryan's algorithm~\cite{bryan} and quadruple numerical precision. 
In the results shown here we have focused on the S waves. 
We begin by looking at the zero temperature spectral functions in the
$\Upsilon$ (vector) channel. Fig.~\ref{fig:zerotemp_mem} 
(left panel) shows the spectral
function at the lowest temperature as a function of the energy. The dotted
lines indicate the ground- and first excited state determined using 
exponential fits to correlators. We observe that there
is very good agreement with the MEM spectral function and that the ground-
and first excited states are clearly determined.
The right panel in Fig.~\ref{fig:zerotemp_mem} shows the temperature dependence in the 
$\Upsilon$ channel with each pane showing two neighboring temperatures. 
The ground state appears to survive to the highest temperature in 
our simulations ($\sim 2T_c$), although the height of the peak is diminished. 
The first excited state is gradually being suppressed
and is no longer discernible at $T/T_c\sim 1.68$ indicating a melting
temperature below $1.68T_c$.
\subsection{Summary}
In this paper we describe recent results from a study of bottomonium at non-zero 
temperature from lattice QCD. The NRQCD formalism offers a number of advantages at 
finite temperature making it a promising line of investigation. We find evidence
of strong temperature dependence in the P waves at $T>T_c$ and indications that while the 1S 
states survive up to $2T_c$ the 2S states are affected much more and appear to 
have melted at temperatures above $1.4T_c$.
%%%%%%%%%%%%%%%%%%%%%%%%%%%%%%%%%%%%%%%%%%%%%%%%%
%% BACKMATTER
%%%%%%%%%%%%%%%%%%%%%%%%%%%%%%%%%%%%%%%%%%%%%%%%
\begin{theacknowledgments}
SR is supported by the Research Executive Agency (REA) of the 
European Union under Grant Agreement number PITN-GA-2009-238353 
(ITN STRONGnet) and the Science Foundation Ireland, Grant No. 11/RFP.1/PHY/3201. 
\end{theacknowledgments}
\bibliographystyle{aipproc}   % if natbib is available

\end{document}